\documentclass[aps,prb,twocolumn,showpacs]{revtex4}

\usepackage{amsmath}
\usepackage{graphicx}
\usepackage{setspace}
\usepackage{bm}
\usepackage{dcolumn}
\usepackage{ulem}

\begin{document}

\title{Infinite series models of flux relaxation and vortex penetration constructed at critical points and their unification}

\author{Rongchao Ma}
\affiliation{Department of Physics, University of Alberta, Edmonton, AB T6G 2E1, Canada}
\email{marongchao@yahoo.com}

\date{\today}

\begin{abstract}
The information about the current-carrying ability of a type-II superconductor can be obtained by studying the flux relaxation and vortex penetration phenomena in the superconductor. In early studies, the infinite series models of the flux relaxation and vortex penetration phenomena were constructed at a vanishing current density and vanishing internal field, respectively. However, this is not the only possibility. Here it is shown that one can reconstruct the theoretical models at the critical points. The new polynomial model of the flux relaxation (vortex penetration) phenomenon was constructed by expanding the vortex activation energy as an infinite series of the current density (internal field) about the critical current density (equilibrium internal field). The unification of the polynomial models was proposed. The inverse model of the flux relaxation (vortex penetration) phenomenon was also constructed by expanding the vortex activation energy as an infinite series of the inverse current density (inverse internal field) about the critical current density (equilibrium internal field). 
\end{abstract}


\maketitle

\section{Introduction}

In the study of a flux relaxation and vortex penetration phenomenon, an important step is to find out an expression of the vortex activation energy. The flux relaxation and vortex penetration can be regarded as the processes of vortex hopping between the adjacent pinning centers.\cite{Ma1,Ma2,Ma3} Therefore, they can be described by the Arrhenius relation \cite{Anderson,Tinkham,Ma2}, which shows that the vortex hopping rate has a strong dependence on the temperature and vortex activation energy. Because the vortex hopping rate can be changed by a current density (or internal field), the activation energy is a function of the current density (or internal field). By proposing a detailed activation energy, one can obtain the corresponding time evolution equation of the current density (or internal field). A number of activation energies were proposed on the basis of different physical considerations. In the flux relaxation process, we have linear activation energy \cite{Anderson}, quadratic activation energy, series activation energy \cite{Ma1,Ma3}, inverse-power activation energy \cite{Feigel'man1,Feigel'man2} and logarithmic activation energy \cite{Zeldov1,Zeldov2}. In the vortex penetration process, we have linear activation energy, quadratic activation energy, and series activation energy.\cite{Ma2,Ma3}

One can see that the flux relaxation and vortex penetration phenomena are currently described with multiple models. The present work is motivated by the following facts:

First, it is now understood that the early developed infinite series models of flux relaxation and vortex penetration \cite{Ma1,Ma2,Ma3} are constructed at zero points (vanishing current density and vanishing internal field). From mathematics, we know that one needs to choose a ``center point'' when doing a Taylor series expansion of a function. This means that it is possible to reformulate the theoretical models by choosing the new center points. The critical current density and equilibrium internal field are important parameters in the application of a superconductor.\cite{Larbalestier} Therefore, we are interested in reconstructing the polynomial models at these critical points.

Second, the inverse-power activation energy \cite{Feigel'man1,Feigel'man2} and logarithmic activation energy \cite{Zeldov1,Zeldov2} are functions of inverse current density. They give good description to the flux relaxation process over the random pinning in the high-$T_c$ superconductors. This led us to consider whether the ``inverse current dependence'' implies a more general physical law that includes the inverse-power activation energy and logarithmic activation energy as special cases. It could be meaningful to extract out this common property and describe it with a unified mathematical equation which can be applied to a wide class of vortex systems.

Finally, the activation energy of a vortex penetration process is currently expressed as a function of the internal field.\cite{Ma2,Ma3} But the activation energy with an inverse internal field dependence is unavailable. Similar to that in the flux relaxation process, therefore, it is desirable to propose an inverse model for the vortex penetration process. The inverse models must be constructed at the critical points because they diverge at vanishing current density (or vanishing internal field).

In this article, we expanded the vortex activation energies at the critical points and developed the new polynomial models and inverse models of vortex dynamics. Surface barrier \cite{Bean} is included in the activation energy, but geometrical barrier \cite{Zeldov3,Gardner} is ignored because it depends on the shape of sample. The information about the anisotropy of the superconductors is included in the activation energy, which is related to the coherence length and penetration depth.

\section{\label{Classification} Classification of vortex activation energy}

A number of vortex activation energies were proposed. Although these activation energies were constructed on the basis of different physical considerations, their mathematical expressions show common features. By carefully checking the existing activation energies, we find that they fall into two classes: 
\\
\\
\textbf{Class A: Polynomial Activation Energy} 

In the flux relaxation process, we have:\\
1. Series activation energy \cite{Ma1,Ma3} $U_r(j)=U_0 - \sum_{l=1}^\infty a_l j^l$, \\
2. Quadratic activation energy \cite{Ma1,Ma3} $U_r(j)=U_0 - a_1 j - a_2 j^2$, \\
3. Linear activation energy \cite{Anderson,Ma1,Ma3} $U_r(j)=U_0 - a_1j$.

These activation energies are decreasing functions of the current density $j$ and are well-defined functions for all $j \in \left[0,j_c\right]$, where $j_c$ is the critical current density.

In the vortex penetration process, we have: \\
1. Series activation energy \cite{Ma2,Ma3} $U_p(B)=U_0 + \sum_{l=1}^\infty c_l B^l$, \\
2. Quadratic activation energy $U_p(B)=U_0 + c_1 B + c_2 B^2$, \\
3. Linear activation energy $U_p(B)=U_0 + c_1 B$. 

These activation energies are increasing functions of the internal field $B$ and are well-defined functions for all $B \in \left[0,B_e\right]$, where $B_e$ is the equilibrium internal field (maximum internal field) of the vortex penetration process. 
\\
\\
\textbf{Class B: Inverse Activation Energy} 

In the flux relaxation process, we have: \\
1. Inverse-power activation energy \cite{Feigel'man1,Feigel'man2} $U_r(j)=U_0(j_c/j)^\mu$, \\
2. Logarithmic activation energy \cite{Zeldov1,Zeldov2} $U_r(j)=U_0 ln(j_c/j)$. 

These activation energies are functions of the inverse current density $1/j$ and diverge at $j=0$. In the vortex penetration process, the inverse activation energy is currently unavailable.

In Class A, the infinite series activation energies are the generalizations of other activation energies. From mathematics, we know that the Taylor series of a function $f(x)$ about a point $x_0$ can be written as
\begin{equation}
\label{TaylorSeries}
f(x)=f[x_0+(x-x_0)]=\sum_{l=0}^\infty a_l (x-x_0)^l,
\end{equation}
where $a_l=f^{(l)}(x_0)/l!$. This shows that, to express a function as a Taylor series, one has to choose a center point $x_0$. By carefully checking the infinite series activation energies in Class A, we see that they are the Taylor series expansions about the point $x_0=0$, more specifically, the expansion of $U_r$ about the point $j=0$ and $U_p$ about the point $B=0$. These series are known as Maclaurin series. However, there is no reason why we limit ourselves to the zero point. We should consider the possibility of constructing the new polynomial models of the flux relaxation and vortex penetration by choosing other center points. The critical points are good candidates.

On the other hand, the activation energies in Class B have inverse current dependence. These activation energies can be regarded as the special models constructed at the critical current density $j_c$. Therefore, we should generalize this ``inverse current dependence'' to an infinite series and apply this method to the vortex penetration phenomenon as well.

\section{Polynomial model}

In this section, we first constructed a new polynomial model of flux relaxation by expanding the activation energy $U_r$ about the critical current density $j_c$. Next, we constructed a new polynomial model of vortex penetration by expanding the activation energy $U_p$ about the equilibrium internal field $B_e$.

\subsection{Polynomial model of flux relaxation phenomenon}

\subsubsection{Current dependence of the activation energy in a flux relaxation process}

In the flux relaxation process, the activation energy $U_r$ is a decreasing function of the current density $j$.\cite{Ma1,Ma3} To construct the polynomial model at the critical current density $j_c$, let us define a normalized current density $\alpha=j/j_c$. We have $\alpha \in [0, 1]$ because $j \in [0, j_c]$.

Referring to Eq.(\ref{TaylorSeries}) and rewriting the activation energy as $U_r[1+(\alpha-1)]$, we see that $U_r$ needs to be expanded as a decreasing function of $(\alpha-1)$ about the point $\alpha=1$. Let us explicitly write out the series expression of $U_r$ as
\begin{equation}
\label{UrPoly}
U_r\left(\alpha\right) = -\sum\limits_{l=1}^\infty a_l \left( \alpha-1 \right)^l,
\end{equation}
where $U_r(1)=0$ ($U_r$ is zero at $j_c$), $a_1=-U'_r(1)$, $a_2=-U''_r(1)/2!$, $\cdots$, $a_l=-U^{(l)}_r(1)/l!$.

With Eq.(\ref{UrPoly}), one can directly go to Subsection \ref{TimeDepe} and derive the time dependence of the current density. Before we do that, let us discuss the pinning potential and the vortex activation energy at a vanishing current density.

In a flux relaxation process, the external field is usually zero. At a vanishing current density, a vortex inside the superconductor is subjected to a pinning force. If the vortex is close to the surface, then it is further subjected to an attractive surface imaging force. This means that the vortex activation energy at the vanishing current density, $U_{r0}$, should includes the pinning potential $U_c$ and the reduction to the activation energy caused by the surface imaging force $U_{im} = (\Phi_0/ 4 \pi \lambda )^2 K_0(2x/\lambda)$. Thus, we have 

\begin{equation}
\label{Ur01}
U_{r0} = U_c - \left( \frac{\Phi_0}{4 \pi \lambda} \right)^2 K_0 \left( \frac{2x}{\lambda} \right).
\end{equation}

On the other hand, $U_{r0}$ can be obtained by putting $\alpha=0$ in Eq.(\ref{UrPoly}), that is,
\begin{equation}
\label{Ur02}
U_{r0} = \sum\limits_{l=1}^\infty (-1)^{l+1} a_l.
\end{equation}

Combing Eq.(\ref{Ur01}) and Eq.(\ref{Ur02}), we have
\begin{equation}
\label{Ur03}
U_c = \sum\limits_{l=1}^\infty (-1)^{l+1} a_l + \left( \frac{\Phi_0}{4 \pi \lambda} \right)^2 K_0 \left( \frac{2x}{\lambda} \right).
\end{equation}

The coefficients $a_l$ can be obtained from experiments (see Eq.(\ref{JPoly})). Doing a measurement at a position away from the surface where $U_{im} = (\Phi_0/ 4 \pi \lambda )^2 K_0(2x/\lambda)=0$, one can calculate the pinning potential $U_c$ using Eq.(\ref{Ur03}).

\subsubsection{\label{TimeDepe} Time dependence of the current density in a flux relaxation process}

To obtain the time dependence of the current density $j(t)$, we still need the time dependence of the activation energy $U_r(t)$. In the flux relaxation process, $U_r(t)$ is an increasing function of the time $t$. An early study \cite{Ma4} has shown that, with logarithmic accuracy, $U_r(t)$ can be written as
\begin{equation}
\label{UrTime}
U_r(t) = k T ln \left(1+\frac{t_i+t}{\tau}\right),
\end{equation}
where $k$ is the Boltzmann constant, $T$ is temperature, $\tau$ is a short time scale parameter, and
\begin{equation}
\label{tiUr}
t_i = \tau \left( e^{U_i/kT}-1 \right)
\end{equation}
is a virtual time interval, during which the activation energy increase from $0$ to the initial value $U_i$.

Combining Eq.(\ref{UrPoly}) and Eq.(\ref{UrTime}), we have
\begin{equation}
\label{UrPoly2}
w_r(t) = \sum\limits_{l=1}^\infty a_l \left( \alpha - 1 \right)^l,
\end{equation}
where
\begin{equation}
\label{wr}
w_r(t) = - k T ln \left(1+\frac{t_i+t}{\tau}\right).
\end{equation}

Inverting Eq.(\ref{UrPoly2}), we have
\begin{equation}
\label{UrPoly3}
\alpha - 1 = \sum\limits_{l=1}^\infty b_l \left[ w_r(t) \right]^l.
\end{equation}

The coefficients $b_l$ in Eq.(\ref{UrPoly3}) are \cite{Ma1}
\begin{equation}
\label{bns}
\begin{aligned}
b_l=& \frac{1}{a_1^l} \frac{1}{l} \sum\limits_{s,t,u \cdots} (-1)^{s+t+u+\cdots} \cdot  \\
    & \frac{l(l+1)\cdots(l-1+s+t+u+\cdots)}{s!t!u!\cdots} \cdot \\
    & \left(\frac{a_2}{a_1}\right)^s \left(\frac{a_3}{a_1}\right)^t \left(\frac{a_4}{a_1}\right)^u \cdots, \\    
\end{aligned}
\end{equation}
where $s+2t+3u+\cdots=l-1$. The inverse coefficients $a_l$ can be obtained by doing a commutation $b_l \leftrightarrow a_l$.

Using the definition $\alpha=j/j_c$, we can rewrite Eq.(\ref{UrPoly3}) as
\begin{equation}
\label{JPoly}
j(t) = j_c \left\{ 1 + \sum\limits_{l=1}^\infty b_l \left[ w_r(t) \right]^l \right\}.
\end{equation}

Eq.(\ref{JPoly}) represents the time dependence of the current density in a flux relaxation process. Putting $t=0$ in Eq.(\ref{JPoly}), we obtain an initial current density
\begin{equation}
\label{JiPoly}
j_i = j_c \left\{ 1 + \sum\limits_{l=1}^\infty b_l \left[ - k T ln \left(1+\frac{t_i}{\tau}\right) \right]^l \right\}.
\end{equation}

Inverting Eq.(\ref{JiPoly}) (or substituting Eq.(\ref{UrPoly}) into Eq.(\ref{tiUr})), we obtain the time parameter $t_i$ expressed in terms of the initial current density $j_i$
\begin{equation}
\label{TiJi2}
t_i = \tau \left\{\prod\limits_{l=1}^\infty e^{-a_l (\alpha_i-1)^l/kT} - 1 \right\},
\end{equation}
where $\alpha_i=j_i/j_c$.

Eq.(\ref{UrPoly}) and Eq.(\ref{JPoly}) are the main formulas of the new polynomial model of the flux relaxation process. The fitting of Eq.(\ref{JPoly}) to the experimental data from a ring shape $Bi_2Sr_2CaCu_2O_{8+x}$ single crystal \cite{Ma5} is shown in Fig.1.

\begin{figure}[htb]
\label{Fig1}
\begin{center}
\includegraphics[height=0.30 \textwidth]{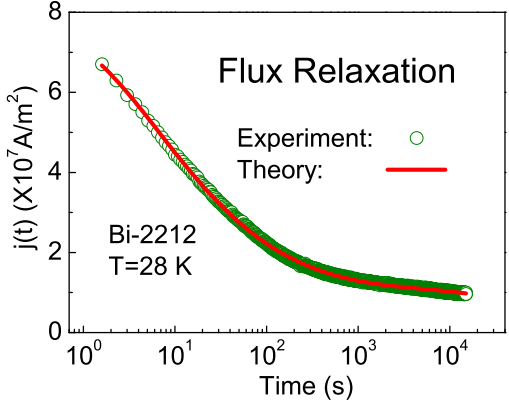}
\caption{ (Color online) Polynomial model of flux relaxation. The scattering points are the time dependence of the persistent current induced at 28 K in a ring shape $Bi_2Sr_2CaCu_2O_{8+x}$ crystal.[Ref.~\onlinecite{Ma5}] The solid black line is the theoretical fit with Eq.(\ref{JPoly}): $ j(t) = j_c \left\{ 1+ \sum_{l=1}^3 b_l [ w_r(t) ]^l \right\} $. The fitting results are:
$j_c=(9.36 \pm 0.03) \times 10^7$, $t_i= 0.79 \pm 0.04$, $\tau=1.18 \pm 0.01$, $b_1=(0.2963 \pm 0.0005)/(28k)$, $b_2=(0.0341 \pm 0.0001)/(28k)^2$, $b_3=(0.0014 \pm 0.0000)/(28k)^3$. }
\end{center}
\end{figure}

\subsubsection{Unified polynomial model of flux relaxation phenomenon}

We have constructed the polynomial model of the flux relaxation phenomenon at the critical current density $j_c$ in Eq.(\ref{UrPoly}) and Eq.(\ref{JPoly}). We have also constructed the polynomial model of the flux relaxation phenomenon at a vanishing current density in Ref.~\onlinecite{Ma1} and ~\onlinecite{Ma3}. Here, we show that these polynomial models can be unified into one theory that can be applied to an arbitrary center point.

Let $j_s \in [0,j_c]$ be an arbitrary center point. Referring to Eq.(\ref{TaylorSeries}), we can expand the activation energy of a flux relaxation process about the point $j_s$ as
\begin{equation}
\label{UrSeries}
U_r(j) = U_r(j_s)+\sum_{l=1}^\infty \tilde{a}_l (j-j_s)^l,
\end{equation}
where $U_r(j_s)$ is the activation energy at the center point $j_s$ and $\tilde{a}_l = U_r^{(l)}(j_s)/l!$. Here, we cannot use the variable $j/j_s$ (see Eq.(\ref{UrPoly})) because it cannot be applied to the zero point $j_s=0$.

Combining Eq.(\ref{UrTime}) and Eq.(\ref{UrSeries}), we have
\begin{equation}
\label{Jfr}
j(t) = j_s + \sum\limits_{l=1}^\infty \tilde{b}_l \left[ \tilde{w}_r(t) \right]^l,
\end{equation}
where
\begin{equation}
\label{wr2}
\tilde{w}_r(t) = k T ln \left(1+\frac{t_i+t}{\tau}\right)-U_r(j_s).
\end{equation}

The relation between $\tilde{a}_l$ and $\tilde{b}_l$ is the same as that of $a_l$ and $b_l$ (see Eq.(\ref{bns})). Using Eq.(\ref{Jfr}), one can analysis the experimental data by choosing an arbitrary center point $j_s \in [0,j_c]$.

Similar to Eq.(\ref{JiPoly}), we have the initial current density
\begin{equation}
\label{JiPoly1}
j_i = j_s + \sum\limits_{l=1}^\infty \tilde{b}_l \left[ k T ln \left(1+\frac{t_i}{\tau}\right)-U_r(j_s) \right]^l
\end{equation}
and the virtual time interval
\begin{equation}
\label{TiJi3}
t_i = \tau \left\{e^{U_r(j_s)/kT}\prod\limits_{l=1}^\infty e^{\tilde{a}_l (j_i-j_s)^l/kT} - 1 \right\}.
\end{equation}

The current density $j_s$ in Eq.(\ref{UrSeries}) and Eq.(\ref{Jfr}) is an arbitrary center point in the interval $[0,j_c]$. If we choose $j_s=0$, then we obtain the polynomial model of the flux relaxation phenomenon constructed at the zero point (see Ref.~\onlinecite{Ma1} and ~\onlinecite{Ma3}). If we choose $j_s=j_c$, then we obtain the polynomial model of the flux relaxation phenomenon constructed at the critical point (see Eq.(\ref{UrPoly}) and Eq.(\ref{JPoly})).

\subsection{Polynomial model of vortex penetration phenomenon}

In the vortex penetration process, it is convenient to use internal field (penetrated field) instead of current density.\cite{Ma2,Ma3,Lin} In this section, therefore, we expanded the activation energy of the vortex penetration process as an infinite series of the internal field.

\subsubsection{Internal field dependence of the activation energy in a vortex penetration process}

In the vortex penetration process, under an applied field $B_a$, the internal field $B$ approaches a maximum value $B_e$ (equilibrium internal field) when the superconductor approaches an equilibrium state. Due to the surface screening effect (or Meissner effect), $B_e$ must be smaller than $B_a$, that is, $B_e<B_a$. (Ref.~\onlinecite{Ma3})

Furthermore, in the vortex penetration process, the activation energy $U_p$ is an increasing function of the internal field $B$. To construct a polynomial model about the equilibrium internal field $B_e$, let us define a normalized internal field $\beta =B/B_e$. We have $\beta \in [0, 1]$ because $B \in [0, B_e]$.

Referring to Eq.(\ref{TaylorSeries}) and rewriting the activation energy as $U_p[1+(\beta-1)]$, we see that $U_p$ needs to be expanded as an increasing function of $(\beta-1)$ about the point $\beta=1$ ($B=B_e$). Let us explicitly write out the series expression of $U_p$ as
\begin{equation}
\label{UpPoly}
U_p(\beta) = U_e + \sum\limits_{l=1}^\infty c_l (\beta-1)^l,
\end{equation}
where $U_e=U_p(1)$, $ c_1=U'_p(1)$, $c_2=U''_p(1)/2!$, $\cdots$, $c_l=U^{(l)}_p(1)/l!$. The parameter $U_e$ is the equilibrium activation energy (at the equilibrium internal field $B_e$) and is also the maximum activation energy in the vortex penetration process under the applied field $B_a$.

Under a vanishing internal field, a vortex inside the superconductor is subjected to the Bean-Livingston \cite{Bean} surface barrier $U_{BL}$ and pinning potential $U_c$. Therefore, the vortex activation energy at the vanishing internal field can be written as 

\begin{equation}
\label{Up01}
U_{p0} = U_{BL} + U_c.
\end{equation}

On the other hand, $U_{p0}$ can also be obtained by putting $\beta=0$ in Eq.(\ref{UpPoly}), that is,
\begin{equation}
\label{Up02}
U_{p0} = U_e + \sum\limits_{l=1}^\infty (-1)^l c_l,
\end{equation}

Combing Eq.(\ref{Up01}) and Eq.(\ref{Up02}), we have
\begin{equation}
\label{Up03}
U_c = U_e + \sum\limits_{l=1}^\infty (-1)^l c_l - U_{BL}.
\end{equation}

The coefficients $U_e$ and $c_l$ can be obtained from experiments (see Eq.(\ref{UpPoly3})). Doing a measurement at a position away from the surface where $U_{BL}=0$, one can calculate the pinning potential using Eq.(\ref{Up03}).

\subsubsection{Time dependence of the internal field in a vortex penetration process}

To obtain the time dependence of the internal field $B(t)$, we still need the time dependence of the activation energy $U_p(t)$. An early study \cite{Ma2} has shown that $U_p(t)$ can be written as
\begin{equation}
\label{UpTime}
U_p(t) = kT ln\left(1 + \frac{t_i+t}{\tau}\right)=kT ln\left(1 + \frac{t_0+t_v+t}{\tau}\right),
\end{equation} 
where $\tau$ is a short time scale parameter and
\begin{subequations}
\begin{eqnarray}
t_i &=& \tau \left( e^{U_i/kT}-1 \right),  \label{tiU} \\
t_0 &=& \tau \left( e^{U_{p0}/kT}-1 \right),  \label{t0U} \\
t_v &=& t_i-t_0 = \tau \left( e^{U_i/kT}-e^{U_{p0}/kT} \right),  \label{tvU}
\end{eqnarray}
\end{subequations}
where $t_0$ is a time parameter equivalent to the potential $U_{p0}$ and $t_v$ is a virtual time parameter, during which the activation energy increase from $U_{p0}$ to the initial value $U_i$.

Combining Eq.(\ref{UpPoly}) and Eq.(\ref{UpTime}), we have
\begin{equation}
\label{UpPoly2}
w_p(t) = \sum\limits_{l=1}^\infty c_l (\beta-1)^l,
\end{equation}
where
\begin{equation}
\label{FunctionWp}
w_p(t) = kT ln\left(1 + \frac{t_0+t_v+t}{\tau}\right) - U_e.
\end{equation}

Inverting Eq.(\ref{UpPoly2}) and using the definition $\beta = B/B_e$, we have
\begin{equation}
\label{UpPoly3}
B(t) = B_e \left\{ 1 + \sum\limits_{l=1}^\infty d_l \left[ w_p(t) \right]^l \right\}.
\end{equation}
 
The coefficients $c_l$ and $d_l$ can be obtained by doing a replacement $a_l \rightarrow c_l$ and $b_l \rightarrow d_l$ in Eq.(\ref{bns}).

Eq.(\ref{UpPoly3}) represents the time dependence of the internal field in a vortex penetration process. Putting $t=0$ in Eq.(\ref{UpPoly3}), we obtain an initial internal field
\begin{equation}
\label{UpPoly4}
B_i = B_e \left\{ 1 + \sum\limits_{l=1}^\infty d_l \left[ kT ln\left(1 + \frac{t_0+t_v}{\tau}\right) - U_e \right]^l \right\}
\end{equation}

If the initial internal field $B_i=0$, then $t_v=0$. Eq.(\ref{UpPoly4}) becomes
\begin{equation}
\label{UpPoly5}
\begin{aligned}
0 &= 1 + \sum\limits_{l=1}^\infty d_l \left[ kT ln\left(1 + \frac{t_0}{\tau}\right) - U_e \right]^l \\
  &= 1 + \sum\limits_{l=1}^\infty d_l \left[ U_{p0} - U_e \right]^l.
\end{aligned}
\end{equation}

Eq.(\ref{UpPoly5}) can also be obtained by inverting Eq.(\ref{Up02}).

Substituting Eq.(\ref{UpPoly}) into Eq.(\ref{tiU}), we have 
\begin{equation}
\label{TiBi2}
t_i = \tau \left\{e^{U_e/kT}\prod\limits_{l=1}^\infty e^{c_l (\beta_i-1)^l/kT} - 1 \right\},
\end{equation}
where $\beta_i=B_i/B_e$.

Substituting Eq.(\ref{UpPoly}) into Eq.(\ref{tvU}), we have
\begin{equation}
\label{TvBi2}
t_v = \tau \left\{e^{U_e/kT}\prod\limits_{l=1}^\infty e^{c_l (\beta_i-1)^l/kT} - e^{U_{p0}/kT} \right\},
\end{equation}

Eq.(\ref{UpPoly}) and Eq.(\ref{UpPoly3}) are the main formulas of the new polynomial model of the vortex penetration process. The fitting of Eq.(\ref{UpPoly3}) to the experimental data from a disk shape $Bi_2Sr_2CaCu_2O_{8+x}$ single crystal \cite{Ma5} is shown in Fig.2.

\begin{figure}[htb]
\label{Fig2}
\begin{center}
\includegraphics[height=0.30 \textwidth]{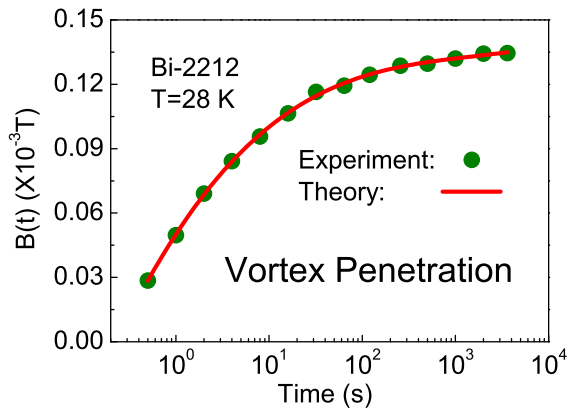}
\caption{ (Color online) Polynomial model of vortex penetration. The scattering points are the time dependence of the internal field induced at 28 K in a disk shape $Bi_2Sr_2CaCu_2O_{8+x}$ crystal.[Ref.~\onlinecite{Ma5}] The solid black line is the theoretical fit with Eq.(\ref{UpPoly3}): $B(t) = B_e \left\{ 1 + \sum_{l=1}^3 d_l [w_p(t)]^l \right\}$. The maximum internal field $B_e = (1.36 \pm 0.01) \times 10^{-4}$ T is determined by the experimental measurements. The fitting results are:
$\tau=0.019 \pm 0.005$, $t_v=0$, $t_0=0.006 \pm 0.001$, $U_e=(12.641 \pm 0.001) \times (28k)$, $d_1=(0.022 \pm 0.000)\times10^{-4}/(28k)$, $d_2=-(0.0049 \pm 0.0008) \times 10^{-4}/(28k)^2$, $d_3=(0.00125 \pm 0.00006)/(28k)^3$. (The $B_e$ should be multiplied by a geometry factor because the field $B(t)$ is measured with a magnetic sensor positioned above the sample. But this does not prevent us from finding out other parameters.) }
\end{center}
\end{figure}

\subsubsection{Unified polynomial model of vortex penetration phenomenon}

We have constructed the polynomial model of vortex penetration phenomenon at the equilibrium internal field $B_e$ in Eq.(\ref{UpPoly}) and Eq.(\ref{UpPoly3}). We have also constructed the polynomial model of vortex penetration phenomenon at a vanishing internal field in Ref.~\onlinecite{Ma2} and ~\onlinecite{Ma3}. Similar to Eq.(\ref{UrSeries}) and Eq.(\ref{Jfr}), let us now construct the unified polynomial model of vortex penetration phenomenon at an arbitrary center point.

Let $B_s \in [0,B_e]$ be an arbitrary center point. Referring to Eq.(\ref{TaylorSeries}), we can expand the activation energy of a vortex penetration process about the point $B_s$ as
\begin{equation}
\label{UpSeries}
U_p(B) = U_p(B_s)+\sum_{l=1}^\infty \tilde{c}_l (B-B_s)^l,
\end{equation}
where $U_p(B_s)$ is the activation energy at the center point $B_s$ and $\tilde{c}_l=U_p^{(l)}(B_s)/l!$. 

Combining Eq.(\ref{UpTime}) and Eq.(\ref{UpSeries}), we have
\begin{equation}
\label{Bvp}
B(t) = B_s + \sum\limits_{l=1}^\infty \tilde{d}_l \left[ \tilde{w}_p(t) \right]^l,
\end{equation}
where
\begin{equation}
\label{vp2}
\tilde{w}_p(t) = k T ln \left(1+\frac{t_i+t}{\tau}\right)-U_p(B_s).
\end{equation}

The relation between $\tilde{c}_l$ and $\tilde{d}_l$ is the same as that of $a_l$ and $b_l$ (see Eq.(\ref{bns})).

Similar to Eq.(\ref{UpPoly4}), we have the initial internal field
\begin{equation}
\label{BiUni}
B_i = B_s + \sum\limits_{l=1}^\infty \tilde{d}_l \left[ k T ln \left(1+\frac{t_i}{\tau}\right)-U_p(B_s) \right]^l,
\end{equation}
the virtual time interval
\begin{equation}
\label{TiUni}
t_i = \tau \left\{e^{U_p(B_s)/kT}\prod\limits_{l=1}^\infty e^{\tilde{c}_l (B_i-B_s)^l/kT} - 1 \right\},
\end{equation}
and
\begin{equation}
\label{TvUni}
t_v = \tau \left\{e^{U_p(B_s)/kT}\prod\limits_{l=1}^\infty e^{\tilde{c}_l (B_i-B_s)^l/kT} - e^{U_{p0}/kT} \right\}.
\end{equation}

In Eq.(\ref{UpSeries}) and Eq.(\ref{Bvp}), if we put $B_s=0$, then we obtain the polynomial model of the vortex penetration phenomenon constructed at the vanishing internal field (see Ref.~\onlinecite{Ma2} and ~\onlinecite{Ma3}). If we put $B_s=B_e$, then we obtain polynomial model of the vortex penetration phenomenon constructed at the critical point (see Eq.(\ref{UpPoly}) and Eq.(\ref{UpPoly3})).

Finally, it should be mentioned that the ``center points'' in the unified polynomial models are not obtained by data fitting, but were chosen by the authors in advance. In other words, the center points are given numbers. This means that, in a vortex penetration process, one needs to know either the equilibrium internal field $B_e$ or equilibrium activation energy $U_e$ in searching for the physical parameters at the critical point (see Fig.2). $B_e$ is determined by the applied field $B_a$, i.e., $B_e \leq B_a$. In a flux relaxation process, however, the activation energy $U_r$ is zero at the critical current density $j_c$. Therefore, one can directly apply Eq.(\ref{JPoly}) or Eq.(\ref{Jfr}) (put $j_s=j_c$) to experimental data to obtain $j_c$.

\section{Inverse model}

In this section, we first presented a generalization to the inverse models of the flux relaxation phenomena discussed in Sec. \ref{Classification}. Next, we applied this method to a vortex penetration phenomenon.

\subsection{Inverse model of flux relaxation phenomenon}

In the inverse model of a flux relaxation process, the activation energy is supposed to be expanded as an infinite series of the inverse current density $1/j$ about the critical current density $j_c$.

\subsubsection{Current dependence of the activation energy in a flux relaxation process}

In the flux relaxation process, the activation energy $U_r$ is a decreasing function of the current density $j$. Therefore, $U_r$ must be an increasing function of $1/j$. To construct the inverse model, let us define a new variable $\lambda=j_c/j$, where $j_c$ is the critical current density. We have	$\lambda \in [1, \infty)$ because $j \in [0,j_c]$ (see Sec. \ref{Discussion}).

Referring to Eq.(\ref{TaylorSeries}) and rewriting the activation energy as $U_r[1+(\lambda-1)]$, we see that $U_r$ needs to be expanded as an increasing function of $(\lambda-1)$ about the point $\lambda=1$ ($j=j_c$). Let us explicitly write out the series expression of $U_r$ as
\begin{equation}
\label{UrInverse}
U_r(\lambda) = \sum\limits_{l=1}^\infty e_l (\lambda-1)^l,
\end{equation}
where $U_r(1)=0$ ($U_r$ is zero at $j_c$), $e_1=U'_r(1)$, $e_2=U''_r(1)/2!$, $\cdots$, $e_l=U^{(l)}_r(1)/l!$.

At a first look, one may doubt the correctness of Eq.(\ref{UrInverse}). To remove this doubt, let us check Eq.(\ref{UrInverse}) with several existing activation energies.

\paragraph{Inverse-power activation energy}\cite{Feigel'man1,Feigel'man2}

\[ \frac{U_0}{\mu} \left(\lambda^\mu-1\right)= U_0 \sum\limits_{l=1}^\infty \frac{(\mu-1)\cdots(\mu-l+1)}{l!} \left(\lambda-1\right)^l. \]

\paragraph{Logarithmic activation energy}\cite{Zeldov1,Zeldov2} 

\[ U_0 ln \lambda = U_0 \sum\limits_{l=1}^\infty \frac{(-1)^{l-1}}{l!} \left( \lambda - 1 \right)^l. \]

\paragraph{Linear activation energy}\cite{Anderson}

\[ U_0 \left( 1-\frac{1}{\lambda} \right)= U_0 \sum\limits_{l=1}^\infty \frac{(-1)^{l+1}}{l!} \left(\lambda-1\right)^l. \]

The other activation energies \cite{Tinkham} can also be checked in the same way, as desired. This confirms that the existing inverse activation energies are the special cases of Eq.(\ref{UrInverse}). Now it is clear that the activation energy $U_r$ can be expressed as an infinite series of the inverse current density $1/j$.

Finally, it should be mentioned that the pinning potential $U_c$ needs to be discussed at a vanishing current density ($j=0$). But in the inverse model, $j=0$ is a singularity of $U_r$ (see Eq.(\ref{UrInverse})). Thus, we do not discuss $U_c$ in the inverse model.

\subsubsection{Time dependence of the current density in a flux relaxation process}

Combining Eq.(\ref{UrTime}) and Eq.(\ref{UrInverse}), we have
\begin{equation}
\label{WvsJ1}
- w_r(t) = \sum\limits_{l=1}^\infty e_l (\lambda-1)^l,
\end{equation}
where $w_r(t)$ is defined by Eq.(\ref{wr}).

Inverting Eq.(\ref{WvsJ1}) and using the definition $\lambda=j_c/j$, we have
\begin{equation}
\label{JrT}
j(t) = j_c \left\{1 + \sum\limits_{l=1}^\infty f_l \left[-w_r(t)\right]^l \right\}^{-1}.
\end{equation}

The coefficients $e_l$ and $f_l$ can be obtained by doing a replacement $a_l \rightarrow e_l$ and $b_l \rightarrow f_l$ in Eq.(\ref{bns}).

Eq.(\ref{JrT}) represents the time dependence of the current density in a flux relaxation process. Putting $t=0$ in Eq.(\ref{JrT}), we obtain an initial current density
\begin{equation}
\label{Js}
j_i = j_c \left\{1 + \sum\limits_{l=1}^\infty f_l \left[k T ln \left(1+\frac{t_i}{\tau}\right)\right]^l \right\}^{-1}.
\end{equation}

Inverting Eq.(\ref{Js}) (or substituting Eq.(\ref{UrInverse}) into Eq.(\ref{tiUr})), we have 
\begin{equation}
\label{TvJR}
t_i = \tau \left\{\prod\limits_{l=1}^\infty e^{e_l (\lambda_i-1)^l/kT} - 1 \right\},
\end{equation}
where $\lambda_i=j_c/j_i$.

Eq.(\ref{UrInverse}) and Eq.(\ref{JrT}) are the main formulas of the inverse model of the flux relaxation process.

\subsection{Inverse model of vortex penetration phenomenon}

In the inverse model of a vortex penetration process, the activation energy is supposed to be expanded as an infinite series of the inverse internal field $1/B$ about the equilibrium internal field $B_e$.

\subsubsection{Internal field dependence of the activation energy in a vortex penetration process}

In the vortex penetration process, the activation energy $U_p$ is an increasing function of the internal field $B$. Thus, $U_p$ must be a decreasing function of $1/B$. To construct the inverse model, let us define a new variable $\sigma=B_e/B$, where $B_e$ is the equilibrium internal field. We have $\sigma \in [1,\infty)$ because $B \in [0,B_e]$ (see Sec. \ref{Discussion}).

Referring to Eq.(\ref{TaylorSeries}) and rewriting the activation energy as $U_p[1+(\sigma-1)]$, we see that $U_p$ needs to be expanded as a decreasing function of $(\sigma-1)$ about the point $\sigma=1$ ($B=B_e$). Let us explicitly write out the series expression of $U_p$ as
\begin{equation}
\label{UpInverse}
U_p(\sigma) = U_e - \sum\limits_{l=1}^\infty g_l (\sigma-1)^l,
\end{equation}
where $U_e=U_p(1)$, $ g_1=-U'_p(1)$, $g_2=-U''_p(1)/2!$, $\cdots$, $g_l=-U^{(l)}_p(1)/l!$. The $U_e$ in Eq.(\ref{UpInverse}) is the equilibrium activation energy corresponding to the equilibrium internal field $B_e$ ($\sigma=1$). This is already discussed in Eq.(\ref{UpPoly}).

The pinning potential $U_c$ needs to be discussed at a vanishing internal field ($B=0$). But $B=0$ is a singularity of $U_p$ (see Eq.(\ref{UpInverse})). Therefore, we do not discuss $U_c$ in the inverse model of the vortex penetration process.

\subsubsection{Time dependence of the internal field in a vortex penetration process}

Combining Eq.(\ref{UpTime}) and Eq.(\ref{UpInverse}), we have
\begin{equation}
\label{WpTau}
-w_p(t) = \sum\limits_{l=1}^\infty g_l (\sigma-1)^l,
\end{equation}
where $w_p(t)$ is defined in Eq.(\ref{FunctionWp}).

Inverting Eq.(\ref{WpTau}) and using the definition $\sigma=B_e/B$, we have
\begin{equation}
\label{BT}
B(t) = B_e \left\{1+\sum\limits_{l=1}^\infty h_l \left[-w_p(t)\right]^l \right\}^{-1}.
\end{equation}

The coefficients $g_l$ and $h_l$ can be obtained by doing a replacement $a_l \rightarrow g_l$ and $b_l \rightarrow h_l$ in Eq.(\ref{bns}).

Eq.(\ref{BT}) represents the time dependence of the internal field in the vortex penetration process. Putting $t=0$ in Eq.(\ref{BT}), we obtain an initial internal field
\begin{equation}
\label{Bi}
B_i = B_e \left\{1+\sum\limits_{l=1}^\infty h_l \left[U_e - kT ln\left(1 + \frac{t_i}{\tau}\right)\right]^l \right\}^{-1}.
\end{equation}

Inverting Eq.(\ref{Bi}) (or substituting Eq.(\ref{UpInverse}) into Eq.(\ref{tiU})), we have
\begin{equation}
\label{TiBi}
t_i = \tau \left\{e^{U_e/kT}\prod\limits_{l=1}^\infty e^{-g_l (\sigma_i-1)^l/kT} - 1 \right\},
\end{equation}
where $\sigma_i=B_e/B_i$.

Eq.(\ref{UpInverse}) and Eq.(\ref{BT}) are the main formulas of the inverse model of the vortex penetration process.

\section{\label{Discussion} Discussion}

On the basis of present work and the early works \cite{Ma1,Ma2,Ma3,Anderson,Feigel'man1,Feigel'man2,Zeldov1,Zeldov2}, we can now summarize the theoretical models of vortex dynamics in Table \ref{table1}.

\begin{table*}
\caption{\label{table1} Vortex activation energies in flux relaxation and vortex penetration process.}
\begin{ruledtabular}
\begin{tabular}{cccc}
&&Flux Relaxation &Vortex Penetration  \\
\hline
I \footnote{Polynomial models constructed about zero point.} &Infinite series [Refs.~\onlinecite{Ma1,Ma2,Ma3}] & $U_a(j) = U_c - \sum\limits_{i=1}^n a_i j^i$,   & $U_b(B) = U_c + \sum\limits_{l=1}^n a_l B^l$,  \\
                                  &Quadratic [Refs.~\onlinecite{Ma1,Ma2,Ma3}] & $U_a(j) = U_c - a_1 j -a_2 j^2$,                & $U_b(B) = U_c + a_l B + a_2 B^2$,   \\
                                  &Linear [Refs.~\onlinecite{Ma1,Ma2,Ma3,Anderson}] & $U_a(j) = U_c - a_1 j$                         & $U_b(B) = U_c + a_l B$             \\
\hline
II \footnote{Polynomial models constructed about critical point. $\alpha=j/j_c$, $\beta=B/B_e$. Expanding the quadratic and linear activation energy, one find that they have the same form as those in I.} &Infinite series & $U_r\left(\alpha\right) = -\sum\limits_{l=1}^\infty a_l \left( \alpha-1 \right)^l$,  & $U_p(\beta) = U_e + \sum\limits_{l=1}^\infty c_l (\beta-1)^l$,  \\
                                     &Quadratic & $U_r\left(\alpha\right) = - a_1 \left( \alpha-1 \right)-a_2 \left( \alpha-1 \right)^2$,  & $U_p(\beta) = U_e +  c_1 (\beta-1) + c_2 (\beta-1)^2$,  \\
                                     &Linear & $U_r\left(\alpha\right) = - a_1 \left( \alpha-1 \right)$  & $U_p(\beta) = U_e + c_1 (\beta-1)$  \\
\hline
III \footnote{Inverse models constructed about critical point. $\lambda=j_c/j$, $\sigma=B_e/B$.} &Infinite series & $U_r(\lambda) = \sum\limits_{l=1}^\infty e_l (\lambda-1)^l$, & $U_p(\sigma) = U_e - \sum\limits_{l=1}^\infty g_l (\sigma-1)^l$  \\
              &Inverse-power [Refs.~\onlinecite{Feigel'man1,Feigel'man2}] & $(U_0/\mu) \left(\lambda^\mu-1\right)$,  & \\
              &Logarithmic [Refs.~\onlinecite{Zeldov1,Zeldov2}] & $U_0 ln \lambda $  & \\
\end{tabular}
\end{ruledtabular}
\end{table*}

\paragraph{Comparison of the polynomial models constructed at zero points and critical points}

The physical parameters of a flux relaxation (vortex penetration) process can be obtained using a polynomial model constructed about either the zero point or the critical point. But from the data fitting, we see that the polynomial models constructed at the critical points have the advantage in determining the physics parameters at the critical points, such as the critical current density $j_c$ in the flux relaxation process, and the equilibrium internal field $B_e$ (or equilibrium activation energy $U_e$) in the vortex penetration process. However, the polynomial models constructed at the zero points \cite{Ma1,Ma2,Ma3} have the advantage in determining the physics parameters at the zero points, such as the pinning potential $U_c$. Thus, for the convenience of calculation, one should choose a proper theoretical model according to the physical parameters that he/she wishes to find out from the experiments.

\paragraph{Comparison of polynomial models and inverse models}

In the polynomial models, the activation energies are well defined functions in the entire domain (Eq.(\ref{UrPoly}) and Eq.(\ref{UpPoly})). Using these formulas, one can calculate various physical parameters in the flux relaxation process and vortex penetration process, such as the pinning potential, critical current density and equilibrium internal field.

In the inverse models, however, the activation energies diverge at the zero points. This defect limits the application of the inverse models. Let us now discuss them separately.

In the inverse model of a flux relaxation process, the activation energy Eq.(\ref{UrInverse}) diverges at $j=0$, although it is a generalization of other inverse current dependent activation energies. Thus, the inverse model cannot describe the vortex behavior at $j=0$ and is only valid at the region close to the critical current density $j_c$. Also, it cannot fit to the experimental data of the entire flux relaxation process, because $j$ reduces from $j_c$ to $0$ with increasing time $t$. This has been confirmed in the fit of experimental data (It is not shown here). Using the inverse model, therefore, one meets difficulty in determining the physical parameters at the zero point, such as the pinning potential $U_c$. Let us now estimate the smallest current density that can be described by the inverse model.

Consider that the maximum activation energy in a flux relaxation process is the pinning potential $U_{r0}$ (see Eq.(\ref{Ur01})). Using Eq.(\ref{UrInverse}), we have,

\[ U_r(\lambda) = \sum\limits_{l=1}^\infty e_l (\lambda-1)^l \leq U_{r0}. \]

Simplifying this inequality, we have

\begin{equation}
\label{MinJ}
j(t) \geq j_c \left( 1 + \sum\limits_{l=1}^\infty f_l U_{r0}^l \right)^{-1}.
\end{equation}

Thus, in the inverse model of a flux relaxation process, Eq.(\ref{JrT}) can only be applied to a system with a current density satisfying the inequality in Eq.(\ref{MinJ}).

Similarly, in the inverse model of a vortex penetration process, the activation energy Eq.(\ref{UpInverse}) diverges at $B=0$. Thus, it cannot describe the vortex behavior at $B=0$ and is only valid at the region close to the equilibrium internal field $B_e$. Also, it cannot fit to the experimental data of the entire vortex penetration process, because $B$ increases from $0$ to $B_e$ with increasing time $t$. Therefore, one meets difficulty in determining the physical parameters at the zero points using the inverse model. Let us now estimate the smallest internal field that can be described by the inverse model.

Consider that the minimum activation energy in a vortex penetration process is the pinning potential $U_{p0}$ (see Eq.(\ref{Up01})). Using Eq.(\ref{UpInverse}), we have

\[ U_p(\sigma) = U_e - \sum\limits_{l=1}^\infty g_l (\sigma-1)^l \geq U_{p0}. \]

Simplifying this inequality, we have

\begin{equation}
\label{MinB}
B(t) \geq B_e \left[ 1+\sum\limits_{l=1}^\infty h_l \left( U_e - U_{p0} \right)^l \right]^{-1}.
\end{equation}

Thus, in the inverse model of a vortex penetration process, Eq.(\ref{BT}) can only be applied to a system with an internal field satisfying the inequality in Eq.(\ref{MinB}).

\paragraph{Surface barrier}

The Bean-Livingston surface barrier \cite{Bean} affects vortex motion. It is usually considered along with the pinning potential at the zero points. In the polynomial models, we directly combined the surface barrier into the activation energies (see Eq.(\ref{Ur01}) and Eq.(\ref{Up01})). In the inverse models, however, the activation energies diverge at the zero points. Thus, we do not consider the surface barrier in the inverse models.

We have ignored the geometrical barrier \cite{Zeldov3,Gardner} in the derivation, because it depends on the sample shape. This effect is significant only for some special geometrical shapes.

\paragraph{Anisotropy of superconductor}

The layered superconductors are strongly anisotropic. This property can be represented by the coherence length and penetration depth. Because the activation energies are the functions of the coherence length and penetration depth, the anisotropy of the superconductors expresses itself in the flux relaxation and vortex penetration process. In this sense, we expect the derived equations to be applicable to both low-$T_c$ and high-$T_c$ superconductors, although we did not explicitly consider the anisotropy of the superconductors in our derivation.

\section{Conclusion}

The polynomial models of flux relaxation and vortex penetration phenomena can be constructed about the critical points. These models are well defined functions in the entire domain. They have advantage in determining the critical current density and equilibrium internal field, comparing with the early proposed polynomial models which were constructed at the zero points. The polynomial models constructed at the zero points and critical points can be unified into one theory. On the other hand, the inverse model of the flux relaxation and vortex penetration phenomena were also constructed at the critical points. The inverse models diverge at the zero points and are only valid at the region close to the critical points. They cannot be used to determine the pinning potential. Therefore, the polynomial models have advantages and give better description to the vortex dynamics.

\end{document}